\documentclass[runningheads]{llncs}
\PassOptionsToPackage{table}{xcolor}  

\usepackage[T1]{fontenc}
\usepackage{graphicx}
\usepackage{booktabs}
\usepackage{multirow}
\usepackage{algorithm}
\usepackage{algorithmic}
\usepackage{amsmath}
\usepackage{subcaption}
\usepackage{tikz}
\usepackage{rotating}
\usetikzlibrary{positioning}
\usepackage{comment}

\usepackage{threeparttable}
\usepackage{lineno}
\usepackage{xcolor}  
\usepackage[colorlinks=true, linkcolor=blue, citecolor=blue, urlcolor=blue]{hyperref}
\usepackage{ulem}
\usepackage{xspace}
\newcommand\modif[1]{\color{black}#1\color{black}\xspace}
\usepackage{orcidlink}
\begin{document}

\author{Tristan Kirscher\inst{1,2}\thanks{Corresponding author: \email{tristan.kirscher@unistra.fr}\\\textbf{Note:} This work has been accepted to MICCAI 2025. The final version will be published in the Lecture Notes in Computer Science (LNCS) series by Springer.} \orcidlink{0009-0004-6646-6548}\and
Sylvain Faisan\inst{2}\orcidlink{0000-0003-3763-9425} 
\and
Xavier Coubez\inst{1}\orcidlink{0000-0002-3791-2009}
\and
Loris Barrier\inst{2}\orcidlink{0009-0000-3394-3601}
\and
Philippe Meyer\inst{1,2}\orcidlink{0000-0002-4505-4737}
}

\authorrunning{T. Kirscher et al.}


\institute{Institut de cancérologie Strasbourg Europe (ICANS), Strasbourg, France \and
ICube Laboratory, CNRS UMR‑7357, University of Strasbourg, France
\\
\email{tristan.kirscher@unistra.fr}}

\title{PSAT: Pediatric Segmentation Approaches via Adult Augmentations and Transfer Learning}
\titlerunning{PSAT: Pediatric Segmentation Approaches}

\maketitle

\begin{abstract}
Pediatric medical imaging presents unique challenges due to significant anatomical and developmental differences compared to adults. Direct application of segmentation models trained on adult data often yields suboptimal performance, particularly for small or rapidly evolving structures. To address these challenges, several strategies leveraging the nnU-Net framework have been proposed, differing along four key axes:
(i) the fingerprint dataset (adult, pediatric, or a combination thereof) from which the Training Plan — including the network architecture—is derived;
(ii) the Learning Set (adult, pediatric, or mixed),
(iii) Data Augmentation parameters, and
(iv) the Transfer \modif{learning} method (fine-tuning versus continual learning).
In this work, we introduce PSAT (Pediatric Segmentation Approaches via Adult Augmentations and Transfer \modif{learning}), a systematic study that investigates the impact of these axes on segmentation performance. We benchmark the derived strategies on two pediatric CT datasets and compare them with state-of-the-art methods, including a commercial radiotherapy solution. 
PSAT highlights key pitfalls and provides actionable insights for improving pediatric segmentation.
Our experiments reveal that a training plan based on an adult fingerprint dataset is misaligned with pediatric anatomy—resulting in significant performance degradation, especially when segmenting fine structures—and that continual learning strategies mitigate institutional shifts, thus enhancing generalization across diverse pediatric datasets.
\modif{The code is available at \url{https://github.com/ICANS-Strasbourg/PSAT}.}

\keywords{Pediatric Segmentation \and Age Bias \and Domain Adaptation \and Transfer Learning}
\end{abstract}
\vspace{1em}

\section{Introduction} \label{sec:intro}

\begin{figure}
    \centering
    \includegraphics[width=.9\columnwidth]{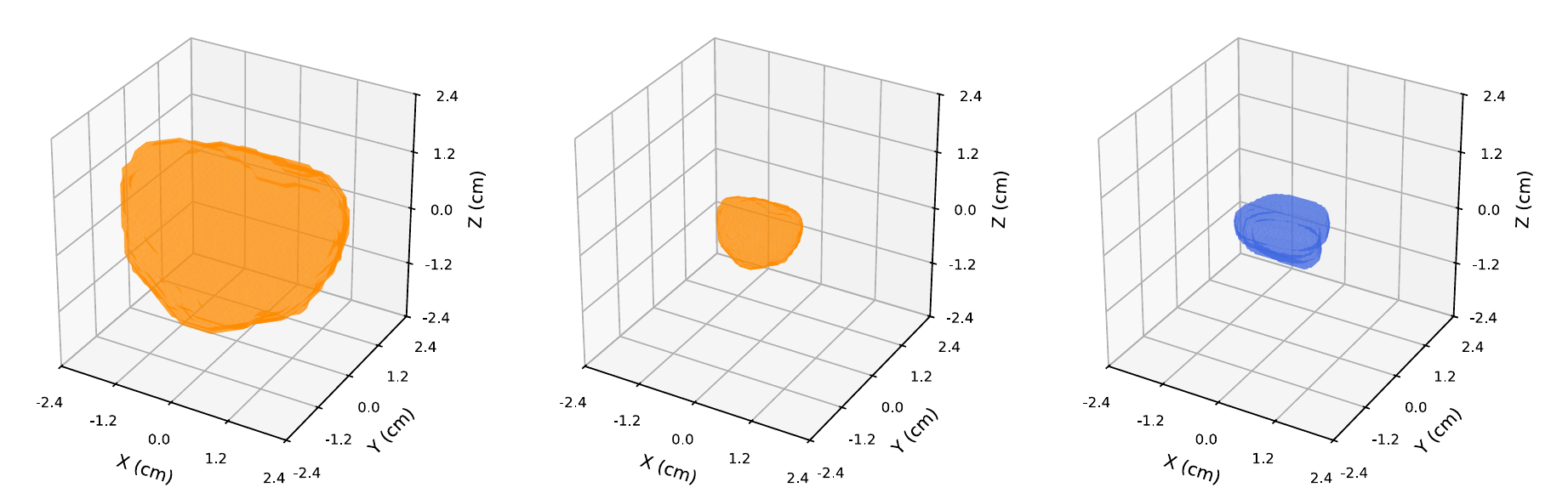}
    \caption{\small Comparison of an adult prostate (orange, 30 cm$^3$), a scaled-down adult prostate with $\sim10\times$ contraction (3 cm$^3$), and a pediatric (2 y.o.) prostate (blue, 3 cm$^3$).}
    \label{fig:roi_scaling}
\end{figure}

Deep learning has revolutionized medical image segmentation, yet its application in pediatric imaging remains particularly challenging. Pediatric scans exhibit marked anatomical differences from adults—including strong organ volume variations (see Fig \ref{fig:roi_scaling}), distinct tissue density profiles, and ongoing developmental changes—that introduce significant domain shifts \cite{somasundaram_deep_2024}. For instance, segmentation accuracy for small organs (e.g., the adrenal gland) can drop from a Dice Similarity Coefficient (DSC) of 0.68 in adults to 0.41 in pediatric cases \cite{chatterjee_children_2024}, with the youngest patients being especially vulnerable \cite{kumar_deep_2024}.

These challenges are further compounded by the scarcity of pediatric-specific annotated data and heterogeneous imaging protocols \cite{sammer_use_2023}. 
\modif{In such settings, transfer learning techniques such as fine-tuning (FT) and continual learning (CL) are commonly used. FT \cite{Tajbakhsh} adapts a pre-trained model to a new task with limited data, while CL \cite{PARISI201954} enables incremental learning without forgetting previously acquired knowledge.} Consequently, many approaches leverage pre-trained adult models within the nnU-Net framework \cite{Isensee2021-xk}, where training plans—including network architecture and preprocessing parameters—are automatically derived from the dataset's imaging fingerprint. Notably, two distinct strategies have been explored to adapt these models for pediatric segmentation. Liu et al. \cite{liu_unlocking_2024} adapt an adult model that was originally configured using an adult fingerprint by integrating both 
\modif{FT and CL}
techniques. In contrast, Chatterjee et al. \cite{chatterjee_children_2024} fine-tune an adult model using a pediatric-specific fingerprint. Both approaches show improvements in segmentation performance; however, they fundamentally differ in how the training plan is derived. It remains uncertain which strategy is the most effective.
Moreover, standard fine-tuning may inadvertently overwrite valuable adult-derived features—a phenomenon known as catastrophic forgetting \cite{gonzalez_lifelong_2023}—which further complicates the adaptation process. Emerging continual learning techniques \cite{liu_unlocking_2024} offer promising alternatives for preserving these features, yet the optimal strategy for pediatric segmentation remains unclear.

To compare these different approaches, we propose PSAT (Pediatric Segmentation Approaches via Adult Augmentations and Transfer \modif{learning}), a framework designed to analyze the influence of different factors on segmentation performance. It considers four key factors: the training plan (P), the learning set composition (S), the data augmentation (A), and the transfer learning strategy (T).
Through comprehensive benchmarking on both public and internal pediatric CT datasets, PSAT yields actionable guidelines for achieving effective pediatric segmentation.

\section{Method}
\label{sec:methods}

\begin{figure}[h]
    \centering
    \includegraphics[width=1\columnwidth]{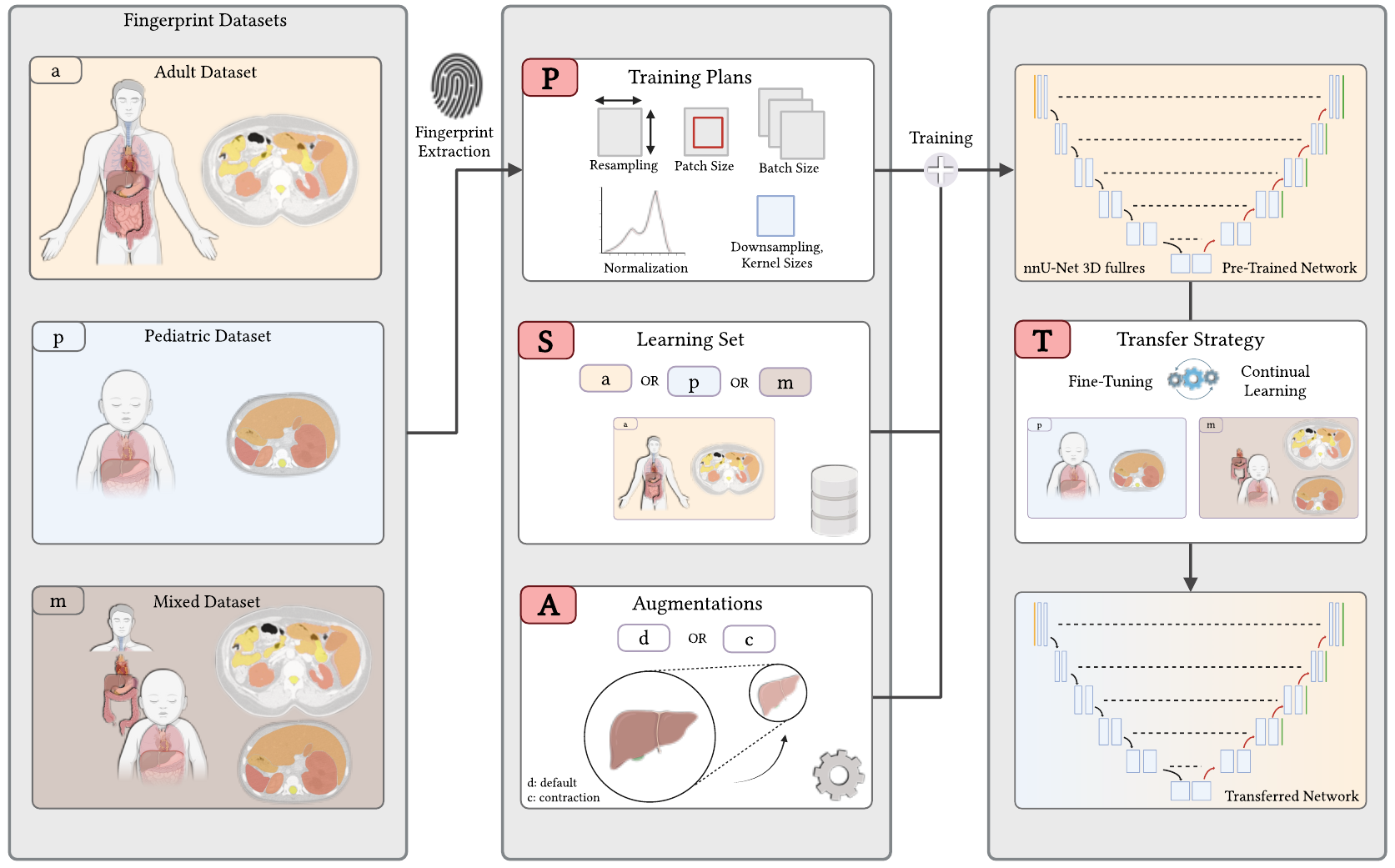}
    \caption{Overview of the PSAT components: (1) Training Plan (fingerprinting via nnU-Net), (2) Learning Set Composition, (3) Data Augmentation, and (4) Transfer Learning Strategies.}
    \label{fig:psat}
\end{figure}

PSAT decomposes the segmentation process into four distinct components (Fig.~\ref{fig:psat}).

\textbf{1. Training Plan:} 
Network configurations and preprocessing parameters (such as resampling protocol and intensity normalization) are derived using the \modif{nnU-Net}
framework from a dataset fingerprint, which includes attributes like median image shape, spacing distribution, and intensity profile.
In our experiments, we consider three nnU-Net 3D fullres configurations: one derived exclusively from an adult dataset ($P_a$), one from a pediatric dataset ($P_p$), and one based on a balanced mixture of adult and pediatric data ($P_m$). 

\textbf{2. Learning Set Composition:}  
We define three learning sets: $S_a$, which comprises solely adult CT scans; $S_p$, which includes only pediatric CT scans; and $S_m$, a combined dataset of adult and pediatric cases.

\textbf{3. Data Augmentation:}  
Two different augmentation strategies can be used during the learning phase. The default one ($A_d$) applies nnU-Net's standard augmentations, including isotropic scaling, rotations, and intensity variations (with mirroring disabled). The contraction-based strategy ($A_c$) differs by allowing isotropic scaling to reduce structure volumes by up to 50\% (vs. 29\% for $A_d$), better mimicking pediatric anatomy. The default augmentation strategy is retained during transfer \modif{learning}.

\textbf{4. Transfer Learning Strategies:}  
We investigate three approaches to adapt pre-trained models to pediatric data. 
\modif{The first one ($T_p$) corresponds to the FT strategy, which}
involves further training models pre-trained on $S_a$  using pediatric data from $S_p$. 
\modif{The second approach ($T_m$) is the CL strategy, which}
employs a mixed rehearsal-based adaptation by integrating both adult and pediatric cases during fine-tuning while maintaining a balanced ratio. Finally, direct inference ($T_o$) utilizes the pre-trained models without additional adaptation.

\section{Results and Discussion}
\label{sec:experiments}

\subsection{Experimental setup}

\paragraph{\textbf{Datasets}}  
We benchmark PSAT on three datasets:

\begin{itemize}
    \item \textbf{Public Pediatric Dataset:} The Pediatric-CT-SEG dataset \cite{jordan_pediatric_2022} consists of 359 pediatric chest-abdomen-pelvis or abdomen-pelvis exams acquired from three CT scanners (ages 0–16), split into training (n=236, mean age: 6.8 ± 4.4, 50\% male), validation (n=59, mean age: 7.7 ± 5.0, 51\% male), and test (n=64, mean age: 6.9 ± 4.6, 50\% male) subsets.
    
    \item \textbf{Public Adult Dataset:} The TotalSegmentator dataset \cite{wasserthal_totalsegmentator_2023} includes 1,082 adult CTs (ages 15–98) with a  wide range of pathologies and institutions. We use the official splits: training (n=937, mean age: 63.4 ± 14.9, 59\% male), validation (n=57, mean age: 64.4 ± 15.2, 60\% male), and test (n=88, mean age: 62.4 ± 16.8, 52\% male).

    \item \textbf{Internal Pediatric Dataset:} This retrospective cohort comprises 50 pediatric CTs (ages 0-16, mean: 7.8 ± 4.7, 54\% male), obtained using varied imaging protocols. The dataset exhibits a significant distributional shift, encompassing a diverse range of cancers—including adrenal, soft tissue, central nervous system, and hematologic tumors.
    The study complied with the ethical rules of the hospital and is registered as \modif{IRB-2025-03} with the hospital institutional review board.
\end{itemize}

The Public Pediatric and the Public Adult Datasets are used for training, transfer learning, and testing, while the Internal Pediatric Dataset is used exclusively for testing to assess the model's robustness to domain shifts.

The organs we aim to segment are those that appear in both public datasets. There are 12 organs in total.

\paragraph{\textbf{Training Procedures}}

Each training follows nnU-Net’s self-configuring pipeline (a batch size of two is maintained across all configurations). Training is conducted on NVIDIA A100 GPUs. During the pre-training phase, models are trained for 1000 epochs using the Adam optimizer with a ``poly'' learning rate decay schedule, starting from an initial learning rate of $10^{-2}$ and decaying to $10^{-5}$ by the end of training.

\modif{For both FT and CL, we use the same ``poly'' learning rate decay schedule as in the pre-training phase, but we perform a grid search over initial learning rates in the interval $[10^{-3},10^{-4}]$ and over the number of epochs in the range $[200,500]$, selecting the combination that yields the highest validation DSC. We use a smaller initial learning rate to preserve pretrained features, which is a common practice in transfer learning (e.g., \cite{sadeghi_adults_2024,Raghu19}). In the CL setting, we employ a rehearsal strategy as described by \cite{gonzalez_lifelong_2023,liu_unlocking_2024}, and perform a grid search over the adult replay ratio in the range [0.25,1]. Overall, we observed that these hyperparameters had a limited impact on final performance, leading us to forego more complex hyperparameter optimization strategies.}

\paragraph{\textbf{PSAT Variants and baselines}} 
We evaluate a selected subset of PSAT configurations.
In the following, a model is considered adult if trained on adult data, pediatric if trained or fine-tuned on pediatric data, or mixed if trained or continuously trained on mixed data.
First, we consider three direct learning approaches, which correspond to the natural way of learning adult ($P_aS_aA_dT_o$), mixed ($P_mS_mA_dT_o$), or pediatric ($P_pS_pA_dT_o$) models, where training plans and learning sets are derived from a single data source. An additional variant employs contraction-based augmentations ($P_aS_aA_cT_o$).
Although nnU-Net is originally designed to generate one model per task/domain (and does not natively support FT), its protocol can be adapted for transfer learning. 
As demonstrated in \cite{chatterjee_children_2024}, an adult model can be pre-trained from a pediatric dataset fingerprint, which is computed from the same pediatric data later used for FT (e.g., $P_pS_aA_dT_p$).
Following this principle, we explore several hybrid learning strategies in which an adult model is trained using either a pediatric dataset fingerprint ($P_pS_aA_cT_o$, $P_pS_aA_dT_o$) or a mixed fingerprint ($P_mS_aA_dT_o$).

We then apply transfer learning to both direct and hybrid models. However, not all possible configurations have been considered: while contraction-based augmentations ($A_c$) influenced the initial model, their impact did not persist so much after transfer learning. To save space, we do not present the transfer \modif{learning} results of models trained with contraction-based augmentations, except for the FT results of the $P_aS_aA_cT_o$ model, referred to as $P_aS_aA_cT_p$. Moreover, we found that transferring from an adult model trained with pediatric training plans was extremely challenging. Due to this limitation, we only present the FT results of one such model, $P_pS_aA_dT_o$, referred to as $P_pS_aA_dT_m$.

We compare PSAT against two baselines: \textit{TotalSegmentator} (TS) v2.4 \cite{wasserthal_totalsegmentator_2023}, an adult-trained model, and the commercial radiotherapy tool \textit{ART-Plan}\texttrademark{} (Therapanacea, France) v2.3.1. It should be noted that ART-Plan\texttrademark{}, while widely used in clinical radiotherapy, is a commercial tool not specifically designed or intended for pediatric applications.

\paragraph{\textbf{Evaluation Metrics and Statistical Analysis}}  
Segmentation performance is assessed using the DSC, defined as $DSC = \frac{2|A \cap B|}{|A|+|B|}$, where $A$ and $B$ denote predicted and ground truth. Due to the non-normal distribution of the DSC data, we applied the Mann–Whitney U test (with $p < 0.05$ indicating significance) to compare models.

\begin{sidewaystable}[htbp]
\centering
\caption{Dice coefficient (\%) comparison across datasets (public adult/public pediatric/internal pediatric).}
\label{tab:results}
\begin{tabular}{lcccccccccc}
\toprule
Trainer & Blad & Duod & Esop & Gall & Kidn & Panc & Pros & Smal & Sple & Stom \\
\midrule
\rowcolor{gray!30}
\multicolumn{11}{c}{\textbf{Baseline}} \\
\midrule
\rowcolor{gray!10} ART-Plan\texttrademark{} & $\cdot$/46/65 & $\cdot$/33/- & $\cdot$/53/59 & * & $\cdot$/79/87 & * & $\cdot$/15/47 & $\cdot$/55/- & $\cdot$/80/88 & $\cdot$/69/80 \\
TS & \textbf{93}/78/85 & 85/47/- & 95/66/61 & \textbf{87}/74/- & \textbf{95}/95/80 & 89/75/49 & \textbf{82}/14/30 & \textbf{92}/62/- & 98/92/87 & 96/86/82 \\
\midrule
\rowcolor{gray!30}
\multicolumn{11}{c}{\textbf{Direct Learning}} \\
\midrule
\rowcolor{gray!10} $P_aS_aA_cT_o$ & 92/80$^{\dagger}$/84 & 85/49$^{\dagger}$/- & 95/66/61 & 86/75$^{\dagger}$/- & 95/95/76 & \textbf{90}/76$^{\dagger}$/46 & 81/24/36 & 92/58/- & 98/93$^{\dagger}$/89 & \textbf{96}/86/83 \\
$P_aS_aA_dT_o$ & 92/66/66 & 85/44/- & 95/64/57 & 84/76$^{\dagger}$/- & 94/95/73 & 90/74/42 & 81/13/14 & 92/55/- & 98/92/88 & 95/85/82 \\
\rowcolor{gray!10} $P_mS_mA_dT_o$ & 92/87$^{\dagger}$/\textbf{89$^{\dagger}$} & 85/75$^{\dagger}$/- & 94/79$^{\dagger}$/\textbf{66$^{\dagger}$} & 85/89$^{\dagger}$/- & 95/97$^{\dagger}$/\textbf{93$^{\dagger}$} & 88/82$^{\dagger}$/\textbf{52$^{\dagger}$} & 80/57$^{\dagger}$/49 & 91/84$^{\dagger}$/- & 98/\textbf{96$^{\dagger}$}/\textbf{91$^{\dagger}$} & 95/93$^{\dagger}$/\textbf{85} \\
$P_pS_pA_dT_o$ & 78/\textbf{90$^{\dagger}$}/84 & 53/\textbf{75$^{\dagger}$}/- & 72/\textbf{80$^{\dagger}$}/64 & 68/\textbf{89$^{\dagger}$}/- & 83/\textbf{97$^{\dagger}$}/91$^{\dagger}$ & 63/\textbf{83$^{\dagger}$}/49 & 58/\textbf{62$^{\dagger}$}/55 & 69/\textbf{84$^{\dagger}$}/- & 92/96$^{\dagger}$/89$^{\dagger}$ & 82/\textbf{94$^{\dagger}$}/83 \\
\midrule
\rowcolor{gray!30}
\multicolumn{11}{c}{\textbf{Hybrid Learning}} \\
\midrule
\rowcolor{gray!10} $P_mS_aA_dT_o$ & 92/72/80 & \textbf{86}/45/- & \textbf{95}/64/57 & 85/76$^{\dagger}$/- & 95/96$^{\dagger}$/75 & 89/74/43 & 82/10/29 & 92/54/- & \textbf{98}/93$^{\dagger}$/88 & 96/86/82 \\
$P_pS_aA_cT_o$ & 89/79$^{\dagger}$/84 & 81/52$^{\dagger}$/- & 92/66/61 & 84/77$^{\dagger}$/- & 93/96$^{\dagger}$/88 & 87/75/40 & 81/13/11 & 87/64/- & 97/94$^{\dagger}$/89 & 94/87$^{\dagger}$/81 \\
\rowcolor{gray!10} $P_pS_aA_dT_o$ & 89/79$^{\dagger}$/82 & 81/50$^{\dagger}$/- & 92/65/61 & 83/77$^{\dagger}$/- & 92/96$^{\dagger}$/85 & 86/73/38 & 81/11/13 & 87/63/- & 97/93$^{\dagger}$/87 & 94/86/80 \\
\midrule
\rowcolor{gray!30}
\multicolumn{11}{c}{\textbf{Transfer Learning}} \\
\midrule
$P_aS_aA_cT_p$ & 90/84$^{\dagger}$/85 & 77/68$^{\dagger}$/- & 84/75$^{\dagger}$/61 & 76/85$^{\dagger}$/- & 83/96$^{\dagger}$/89 & 84/78$^{\dagger}$/50$^{\dagger}$ & 0/0/0 & 81/80$^{\dagger}$/- & 95/95$^{\dagger}$/85 & 91/90$^{\dagger}$/78 \\
\rowcolor{gray!10} $P_aS_aA_dT_m$ & 92/85$^{\dagger}$/87$^{\dagger}$ & 83/71$^{\dagger}$/- & 94/77$^{\dagger}$/64 & 81/87$^{\dagger}$/- & 93/97$^{\dagger}$/93$^{\dagger}$ & 88/80$^{\dagger}$/51$^{\dagger}$ & 80/51$^{\dagger}$/\textbf{57} & 89/81$^{\dagger}$/- & 98/96$^{\dagger}$/90$^{\dagger}$ & 94/91$^{\dagger}$/83 \\
$P_aS_aA_dT_p$ & 86/88$^{\dagger}$/83 & 75/73$^{\dagger}$/- & 82/77$^{\dagger}$/61 & 78/87$^{\dagger}$/- & 90/97$^{\dagger}$/93$^{\dagger}$ & 84/81$^{\dagger}$/50$^{\dagger}$ & 0/0/0 & 83/82$^{\dagger}$/- & 97/96$^{\dagger}$/84 & 92/92$^{\dagger}$/81 \\
$P_mS_aA_dT_m$ & 92/87$^{\dagger}$/88$^{\dagger}$ & 83/71$^{\dagger}$/- & 93/78$^{\dagger}$/65$^{\dagger}$ & 81/86$^{\dagger}$/- & 92/97$^{\dagger}$/93$^{\dagger}$ & 88/81$^{\dagger}$/50$^{\dagger}$ & 81/58$^{\dagger}$/56 & 89/82$^{\dagger}$/- & 98/96$^{\dagger}$/90$^{\dagger}$ & 94/92$^{\dagger}$/83 \\
\rowcolor{gray!10} $P_mS_aA_dT_p$ & 89/87$^{\dagger}$/86 & 73/73$^{\dagger}$/- & 88/79$^{\dagger}$/64 & 79/87$^{\dagger}$/- & 89/97$^{\dagger}$/93$^{\dagger}$ & 85/81$^{\dagger}$/50$^{\dagger}$ & 76/60$^{\dagger}$/57 & 83/82$^{\dagger}$/- & 97/96$^{\dagger}$/89$^{\dagger}$ & 92/92$^{\dagger}$/82 \\
$P_pS_aA_dT_p$ & 74/76/73 & 53/45/- & 35/40/29 & 49/62/- & 8/24/14 & 52/58/39 & 0/0/0 & 72/68$^{\dagger}$/- & 69/84/77 & 83/81/76 \\
\bottomrule
\end{tabular}\\
\begin{tablenotes}
\footnotesize
\footnotesize \item \textbf{Bold} marks the best DSC for each ROI/dataset. ROI names are abbreviated: Blad (bladder), Duod (duodenum), Esop (esophagus), Gall (Gallbladder), Kidn (left kidney), Panc (pancreas), Pros (prostate), Smal (small intestine), Sple (spleen), Stom (stomach).
\item $^{\dagger}$ indicates a statistically significant improvement over the best performing baseline ($p < 0.05$). “-” indicates that the physician reference is not available and “*” that the model does not segment this ROI. ART-Plan\texttrademark{} was not tested on adult data ($\cdot$).
\end{tablenotes}
\end{sidewaystable}

\subsection{Analysis of Results} \label{sec:results}

Table~\ref{tab:results} reports the DSC for multiple regions-of-interest (ROIs) across three test sets (public adult/public pediatric/internal pediatric). Liver segmentation (DSC > 90\%) remained robust across all configurations and is omitted for brevity. Only left kidney results are shown, as right kidney had similar performance.

\paragraph{\textbf{Adult Segmentation Performance}}

On the adult test set, both direct and hybrid learning models—trained on adult data ($S_a$) or on a mixed dataset ($S_m$)—achieve performance comparable to the state-of-the-art TS, with DSC differences of only 1–2 percentage points. In particular, the $P_aS_aA_dT_o$ configuration yields nearly identical results to TS, as expected given the shared imaging fingerprint and training set.

Contraction-based augmentations ($A_c$) do not degrade performance on adult data. In contrast, employing pediatric-specific training plans ($P_p$) on the adult set (e.g., $P_pS_aA_dT_o$) results in a modest performance decline (e.g., bladder DSC decreased from 92\% to 89\%, duodenum from 85\% to 81\%, and small intestine from 92\% to 87\%).

Fine-tuning models on pediatric data ($T_p$) exhibits clear signs of catastrophic forgetting. For example, the $P_aS_aA_dT_p$ configuration shows a 13-point drop for the esophagus, a 9-point decrease for the gallbladder, and, most notably, a complete loss of prostate segmentation (DSC = 0\%)—indicating the loss of adult-derived representations. In contrast, CL strategies—such as $P_aS_aA_dT_m$ and $P_mS_aA_dT_m$ —maintain performance levels comparable to those observed with direct learning.

\paragraph{\textbf{Adult Models for Pediatric Segmentation}}
Adult-trained models are evaluated on two pediatric datasets that are unseen during training. Models trained exclusively on adult data ($S_a$) exhibit significantly lower segmentation performance on pediatric cases—a finding consistent with previous reports \cite{somasundaram_deep_2024,chatterjee_children_2024,kumar_deep_2024}.

Incorporating contraction-based augmentations ($A_c$) leads to substantial improvements. For example, the $P_aS_aA_cT_o$ configuration increases the bladder DSC from 66\% to 80\% on the public pediatric dataset and from 66\% to 84\% on the internal pediatric dataset.

Models employing pediatric-specific training plans ($P_p$) achieve higher DSC values on the public pediatric dataset compared to their adult-plan counterparts. For instance, $P_pS_aA_dT_o$ yields improvements of +13 DSC on the bladder, +6 on the duodenum, and +8 on the small intestine relative to $P_aS_aA_dT_o$. 
Nonetheless, the benefits of pediatric-specific training plans are inconsistent across structures for the internal pediatric dataset.
One possible explanation is that the pediatric training plan is inherently tailored to the public pediatric dataset. In contrast, the internal pediatric dataset remains truly unseen.

Finally, TS performs slightly better in pediatrics than the $P_aS_aA_dT_o$ configuration, possibly due to its segmentation of 117 structures, which may enable it to capture spatial dependencies more effectively.

\paragraph{\textbf{Pediatric and mixed Models on the Public Pediatric Dataset}}
All pediatric and mixed models have been exposed to images from the public pediatric dataset during training, transfer learning establishing an intra-institutional setting.

The direct learning pediatric model ($P_pS_pA_dT_o$) achieves the best overall performance, showing substantial improvements over both ART-Plan\texttrademark{} and TS (e.g., prostate +48 DSC, duodenum +28 DSC, gallbladder +25 DSC, esophagus +24 DSC). The direct learning model trained on a mixed learning set ($P_mS_mA_dT_o$) exhibits slightly lower performance than $P_pS_pA_dT_o$, yet consistently outperforms the baselines in nearly all ROIs with statistically significant gains.

FT of an adult model using pediatric-specific training plans produces poor results, likely due to optimization challenges. In our experiments, the model struggles to transition from the adult domain to a pediatric-appropriate minimum, essentially failing to converge when initialized with an adult model. This observation is consistent with \cite{chatterjee_children_2024}, where the authors pre-trained a pediatric network on adult data for \textit{only} 100 epochs before conducting an extensive 4,000-epoch FT phase, ultimately yielding negligible differences between direct learning and FT.

Apart from pediatric-specific plans, transfer learning approaches—whether via FT ($T_p$) or CL ($T_m$)—yield significant gains over the baselines. However, FT from an adult plan ($P_a$) results in catastrophic forgetting of the prostate (DSC = 0). Within this intra-institutional setting, we observed that $P_pS_pA_dT_o$ slightly outperforms $P_mS_mA_dT_o$, suggesting that incorporating adult image information may, to some extent, hinder pediatric performance. This aligns with the observation that FT slightly outperforms CL, except for the prostate when using an adult plan. Overall, the best transfer \modif{learning} appears to be achieved with mixed training plans combined with FT (i.e., $P_mS_aA_dT_p$).

\paragraph{\textbf{Pediatric and mixed Models on the Internal Pediatric Dataset}}
This section examines model generalization to unseen institutions, where pediatric testing images come from a different dataset than those used for training or transfer \modif{learning}. In this inter-institutional scenario, segmentation performance degrades significantly compared to results with public data, likely due to institutional shifts.

Although the direct pediatric model ($P_pS_pA_dT_o$) still outperforms the baseline (e.g., prostate +25 DSC, kidney +4 DSC), the mixed model ($P_mS_mA_dT_o$) now achieves the best overall performance, suggesting that integrating adult image information helps bridge the domain gap. 
\modif{Consistently,}
CL emerges as the optimal transfer \modif{learning} strategy, yielding slightly better performance than FT. This  aligns with \cite{gonzalez_lifelong_2023}, which shows that CL can reduce institutional shift—a benefit that holds true in pediatric segmentation.
As previously observed, FT from an adult plan ($P_a$) leads to catastrophic forgetting (e.g., DSC = 0 for the prostate), establishing that $P_mS_aA_dT_m$ is the most effective transfer \modif{learning} strategy.

\section{Conclusion}
In this work, we introduced PSAT, a systematic study of pediatric segmentation that leverages adult data through diverse training plans, augmentation strategies, and transfer \modif{learning} techniques. Our study highlights \modif{four} key findings:
\begin{enumerate}
    \item In intra-institutional settings, fine-tuning (FT) generally outperforms continual learning (CL); however, when addressing inter-institutional domain shifts, CL emerges as the more robust strategy.
    \item While overlooked in the literature, training plan selection is critical. Transferring from adult-specific plans is particularly risky for certain structures (e.g., the prostate), while pediatric-specific plans can hinder tuning. A mixed training plan—leveraging adult data during pre-training and adapting to pediatric data during transfer \modif{learning}—appears to be the optimal compromise.
    \item Applying contraction-based augmentations on adult data to mimic pediatric organ sizes significantly enhances generalization. Future work should explore organ-specific augmentation strategies to further improve performance.
    \item \modif{Direct learning seems to yield the highest performance; however, in} pediatric 
    \modif{inter-institutional scenarios, CL ($P_mS_aA_dT_m$) performs only marginally below direct learning ($P_mS_mA_dT_o$); similarly, in} pediatric 
    \modif{intra-institutional settings, FT ($P_mS_aA_dT_p$) closely matches direct learning} ($P_pS_pA_dT_o$). \modif{The practical advantage of transfer learning-based strategies lies in their efficiency: when a pretrained adult model is available, CL takes about 10 hours and FT about 2.5 hours, compared to about 25 hours for full retraining. Note that CL requires access to the full adult pretraining dataset and incurs higher computational cost than FT.}
\end{enumerate}

\modif{
\noindent \textbf{Acknowledgments.} \hspace{0.005cm} This work \hspace{0.005cm} of the Interdisciplinary \hspace{0.005cm} Thematic \hspace{0.005cm}Institute HealthTech, as part of the ITI 2021-2028 program of the University of Strasbourg, CNRS and Inserm, was partially supported by IdEx Unistra (ANR-10-IDEX-0002) and SFRI (STRAT’US project, ANR-20-SFRI-0012) under the framework of the French Investments for the Future Program. The authors would like to acknowledge the High Performance Computing Center of the University of Strasbourg for supporting this work by providing scientific support and access to computing resources. Part of the computing resources were funded by the Equipex Equip@Meso project (Programme Investissements d'Avenir) and the CPER Alsacalcul/Big Data. \\

\noindent \textbf{Disclosure of Interests.} The authors have no competing interests to declare that are relevant to the content of this article.
}

\bibliographystyle{splncs04}
\bibliography{biblio}

\end{document}